\documentclass[prl,twocolumn,showpacs,floats]{revtex4}

\usepackage{graphicx}

\begin{document}

\title{Common Features in Electronic Structure of the Fe-Based Layered
Superconductors from Photoemission Spectroscopy}
\author{Xiaowen Jia$^{1}$, Haiyun Liu$^{1}$, Wentao Zhang$^{1}$, Lin Zhao$^{1}$, Jianqiao Meng$^{1}$,
Guodong  Liu$^{1}$, Xiaoli Dong$^{1}$, G. F. Chen$^{1}$, J. L.
Luo$^{1}$, N. L. Wang$^{1}$, Z. A. Ren$^{1}$,  Wei Yi$^{1}$, Jie
Yang$^{1}$, Wei Lu$^{1}$, G. C. Che$^{1}$, G. Wu$^{2}$, R. H.
Liu$^{2}$, X. H. Chen$^{2}$, Guiling Wang$^{1}$, Yong Zhou$^{1}$,
Yong Zhu$^{3}$, Xiaoyang Wang$^{3}$, Zhongxian Zhao$^{1}$, Zuyan
Xu$^{1}$, Chuangtian Chen$^{3}$, X. J. Zhou$^{1,*}$}

\affiliation{
\\$^{1}$Beijing National Laboratory for Condensed
Matter Physics, Institute of Physics, Chinese Academy of Sciences,
Beijing 100190, China
\\$^{2}$Hefei National Laboratory for Physical Sciences at Microscale and Department of Physics,
University of Science and Technology of China, Hefei, Anhui 230026,
China
\\$^{3}$Technical Institute of Physics and Chemistry, Chinese Academy of Sciences, Beijing 100190, China
}
\date{June 2, 2008}
%
%

\begin{abstract}

High resolution photoemission measurements have been carried out on
non-superconducting  LaOFeAs parent compound and various
superconducting R(O$_{1-x}$F$_x$)FeAs (R=La, Ce and Pr) compounds.
We found that the parent LaOFeAs compound shows a metallic
character. Through extensive measurements, we have identified
several common features in the electronic structure of these
Fe-based compounds: (1). 0.2 eV feature in the valence band; (2). A
universal 13$\sim$16 meV feature;  (3). A clear Fermi cutoff showing
zero leading-edge shift in the superconducting state; (4). Lack of
superconducting coherence peak(s); (5). Near E$_F$ spectral weight
suppression with decreasing temperature. These universal features
can provide important information about band structure,
superconducting gap and pseudogap in these Fe-based materials.

\end{abstract}


\maketitle

The discovery of superconductivity in Fe-based
oxypnictides\cite{Kamihara,GFChenLa, HHWen,
XHChen43K,GFChenCe,ZARenNd,ZARenPr,ZARenSm,XHPhaseDiagram}  has
generated a great interest because it represents the second class of
``high temperature superconductors" after the first one in the
cuprates\cite{Bednorz}.  The parent compound of these Fe-based
superconductors has been found to exhibit a spin density
wave(SDW)-like transition near 150 K and a possible
antiferromagnetic ground
state\cite{DongSDW,XHPhaseDiagram,PCDai,McGuireNeutron}. Doping
charge carriers into the system induces superconductivity at
appropriate doping levels.  These behaviors appear to be similar to
those in cuprate superconductors. Some important questions to ask
include: (1). Whether these materials can be categorized into strong
correlated electron systems; (2). Whether the mechanism of
superconductivity is conventional or exotic; (3).Whether the normal
state is anomalous or whether there is a pseudogap in the normal
state as found in the cuprate superconductors\cite{PseudogapReview}.
Photoemission spectroscopy, as a powerful tool to  directly measure
the electronic structure and energy gap, can shed important light on
these issues\cite{ARPESReview,HWOu,Shin,Takahashi,HYLiu}.

\begin{figure}[tbp]
\begin{center}
\includegraphics[width=0.85\columnwidth,angle=0]{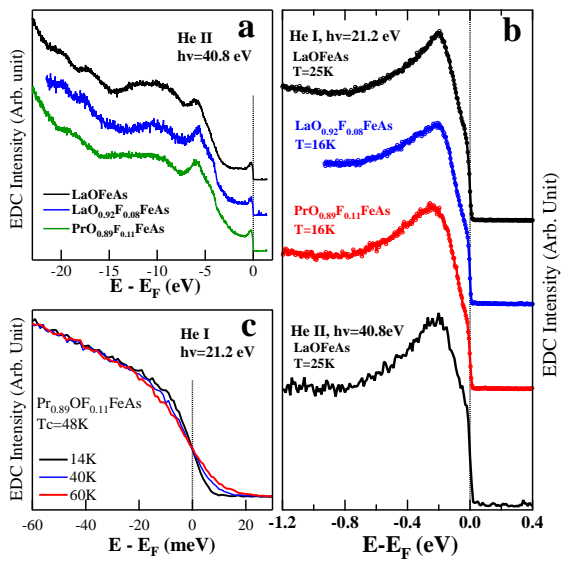}
\end{center}
\caption{Photoemission spectra of R(O$_{1-x}$F$_x$)FeAs (R=La and
Pr) samples measured using 21.2 eV and 40.8 eV photon energies from
the Helium lamp. (a). Large energy range photoemission spectra of
LaOFeAs, LaO$_{0.92}$F$_{0.08}$FeAs and PrO$_{0.89}$F$_{0.11}$FeAs
measured using 40.8 eV photon energy from Helium lamp.  (b). Valence
band of LaOFeAs, LaO$_{0.92}$F$_{0.08}$FeAs and
PrO$_{0.89}$F$_{0.11}$FeAs measured using 21.2 eV and 40.8 eV photon
energies from Helium lamp. (3). Temperature dependence of near-E$_F$
spectra for the PrO$_{0.89}$F$_{0.11}$FeAs sample measured using
21.2 eV photon energy with an energy resolution of 6.5 meV. }
\end{figure}

In this paper, we report high resolution photoemission measurements
on various R(O$_{1-x}$F$_x$)FeAs compounds (R=La, Ce and Pr). We
have found that the parent LaOFeAs compound shows a metallic
behavior which is distinct from the antiferromagnetic insulator in
the parent compound of cuprate superconductors.  Through extensive
measurements, we have identified several common features in the
electronic structure of the Fe-based compounds: (1). 0.2 eV feature
in the valence band; (2). A universal 13$\sim$16 meV feature;  (3).
A clear Fermi cutoff showing zero leading-edge shift in the
superconducting state; (4). Lack of superconducting coherence
peak(s); (5). Near E$_F$ spectral weight suppression with decreasing
temperature. These universal features can provide important
information about band structure, superconducting gap and pseudogap
in these Fe-based compounds.

The photoemission measurements have been carried out on our
newly-developed system using both Vacuum Ultraviolet (VUV) laser
 and Helium discharge lamp as light
sources, equipped with Scienta R4000 electron energy
analyzer\cite{GDLiu,HYLiu}. The laser photon energy (hv) is 6.994 eV
and the spot size is less than 0.2 mm. For the laser measurements,
the energy resolution was set at 1.0 meV. The Helium lamp can
provide two photon energies at 21.218 eV (Helium I resonance line)
and 40.813 eV (Helium II resonance line). The energy resolution for
the 21.218 eV and 40.813 eV valence band measurements was set at
12.5$\sim$20 meV. Polycrystalline R(O$_{1-x}$F$_x$)FeAs (R=La, Ce
and Pr) samples with various dopings (x, nominal composition)  are
prepared by solid state reaction
method\cite{DongSDW,GFChenCe,ZARenPr}. The LaOFeAs sample is not
superconducting but with a possible SDW transition at $\sim$150
K\cite{DongSDW}, while the LaO$_{0.92}$F$_{0.08}$FeAs,
CeO$_{0.88}$F$_{0.12}$FeAs and PrO$_{0.89}$F$_{0.11}$FeAs samples
are superconducting with T$_c$ at 26 K, 38 K and 48 K,
respectively\cite{DongSDW,GFChenCe,ZARenPr}. To get clean surface
and avoid sample aging effect\cite{HYLiu}, all the laser
photoemission data presented in the work were measured on a fresh
sample surface within 3$\sim$4 hours after {\it in situ} fracturing
in vacuum with a base pressure better than 5$\times$10$^{-11}$ Torr.

\begin{figure}[tbp]
\begin{center}
\includegraphics[width=0.85\columnwidth,angle=0]{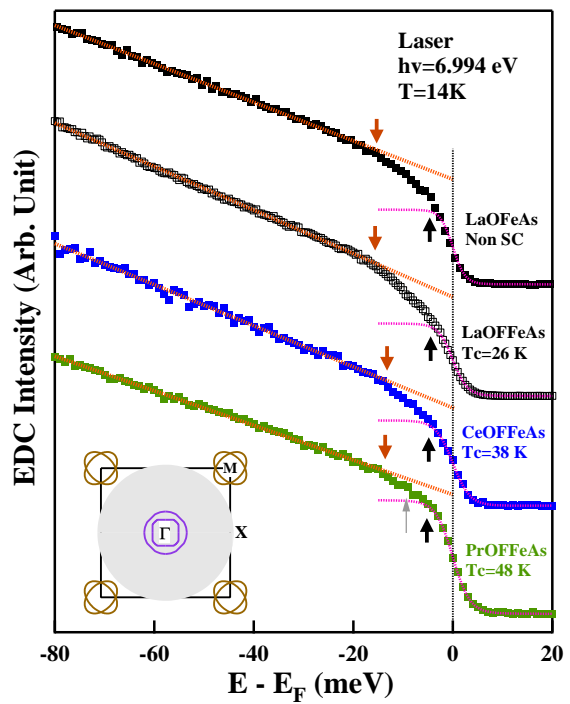}
\end{center}
\caption{Photoemission spectra of LaOFeAs,
LaO$_{0.92}$F$_{0.08}$FeAs, CeO$_{0.88}$F$_{0.12}$FeAs and
PrO$_{0.89}$F$_{0.11}$FeAs samples taken with laser (6.994 eV) at 14
K. The high binding energy part between 20 and 80 meV shows linear
behavior, as indicated by the fitted dashed lines. The curves start
to deviate from the linear line at 13$\sim$16 meV, as indicated by
the down arrows. Upon approaching the Fermi level, another drop
occurs at $\sim$4 meV, as marked by the black up arrows. The spectra
near the Fermi level can be well fitted using Fermi-Dirac
distribution function, as indicated by the purple dotted curves. The
inset shows a schematic Brillouin zone and the calculated Fermi
surface\cite{DongSDW,WorkFNote}. The momentum area that can be
covered by angle-integrated laser photoemission is marked as a
shaded region. }
\end{figure}

Fig. 1 shows photoemission spectra of R(O$_{1-x}$F$_x$)FeAs (R=La
and Pr) samples measured using different photon energies of the
Helium lamp. Over a large energy range (Fig. 1a), these samples show
similar photoemission spectra with two main peaks near 0.2 eV and
$\sim$6 eV, one broad feature between 7$\sim$14 eV, and some weak
peaks at higher binding energy\cite{HWOu,Takahashi,VBNote}. The
valence band shows mainly a pronounced peak at 0.2 eV (Fig. 1b).
Different photon energies give similar spectra as seen from the 21.2
eV and 40.8 eV measurements on the LaOFeAs sample. Different samples
show similar spectra and there is no dramatic spectral change
between the undoped LaOFeAs and doped LaO$_{0.92}$F$_{0.08}$FeAs
samples. Particularly, the parent LaOFeAs compound  shows a clear
Fermi cutoff (Figs. 1 and 2), indicating its metallic nature. This
is distinct from the cuprates where the parent compound is an
antiferromagnetic insulator\cite{ARPESReview}.

\begin{figure}[tbp]
\begin{center}
\includegraphics[width=0.90\columnwidth,angle=0]{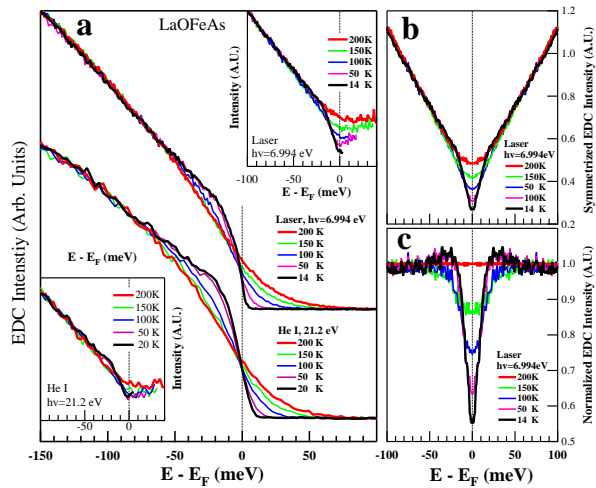}
\end{center}
\caption{(a). Temperature dependence of the photoemission spectra on
the LaOFeAs sample measured by laser and 21.2 eV photon energy from
the helium lamp. The upright and bottom-left insets show the laser
data and helium lamp data, respectively, with the Fermi-Dirac
distribution function removed. (b). Symmetrized spectra from laser
data in (a) with respect to the Fermi level. (c). Symmetrized laser
spectra divided by the the data at 200 K. To avoid statistical
noise, the 200 K data is fitted with a polynomial and used for the
normalization. }
\end{figure}

Fig. 2 shows photoemission spectra of R(O$_{1-x}$F$_x$)FeAs (R=La,
Ce and Pr) measured using laser at 14 K. The LaOFeAs parent compound
and the superconducting samples exhibit similar low energy spectra
where two obvious kink features can be identified. One is at higher
binding energy where the spectrum starts to deviate from the linear
behavior, $\sim$16 meV for LaOFeAs and LaO$_{0.92}$F$_{0.08}$FeAs
and $\sim$13 meV for CeO$_{0.88}$F$_{0.12}$FeAs and
PrO$_{0.89}$F$_{0.11}$FeAs\cite{NotePosition}. The other is at lower
binding energy near 4 meV that is due to the Fermi function cutoff.
The 13$\sim$16 meV feature is robust because it is present in both
undoped sample and doped samples, in different superconducting
materials (R=La, Ce, Pr and Sm\cite{HYLiu}),  and seen in both laser
photoemission data and high resolution helium lamp measured
data\cite{HYLiu}.

The high resolution low temperature data in Fig. 2 provide a good
opportunity to examine the superconducting gap in these Fe-based
superconductors.  Generally speaking, in the momentum-integrated
photoemission spectrum, it is straightforward to determine one
superconducting gap on a single Fermi surface by following the
leading-edge shift, as exemplified in superconducting
diamond\cite{DiamondGap}.  In a system with multiple superconducting
gaps on different Fermi surface sheets, the leading edge position is
mainly dictated by the minimum superconducting gap on a Fermi
surface sheet; the larger energy gaps on the other Fermi surface
sheets will show up as additional features in the spectrum at higher
binding energy, as demonstrated in a two-gap system like
MgB$_2$\cite{MgB2Gap}.  However, the situation may get complicated
if a superconductor has multiple gaps involving unconventional
pairing symmetry and/or does not show clear superconducting
coherence peaks.

\begin{figure}[tbp]
\begin{center}
\includegraphics[width=0.90\columnwidth,angle=0]{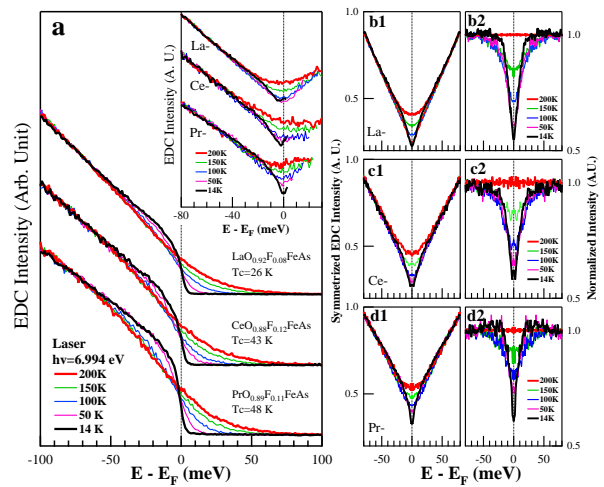}
\end{center}
\caption{(a). Temperature dependence of the laser photoemission
spectra on the La(O$_{0.92}$F$_{0.08}$)FeAs,
CeO$_{0.88}$F$_{0.12}$FeAs and PrO$_{0.89}$F$_{0.11}$FeAs samples.
The upright inset shows the corresponding data but with the
Fermi-Dirac distribution function removed. (b1),(c1) and (d1) show
symmetrized laser spectra from (a) with respect to the Fermi level.
(b2),(c2) and (d2) show corresponding symmetrized spectra divided by
the 200 K data . To avoid statistical noise, the 200 K data is
fitted with a polynomial and used for the normalization. }
\end{figure}

The Fe-based compounds exhibit unusual superconducting behaviors in
two aspects. The first is the lack of superconducting coherence
peaks in the measured photoemission spectra for all these
R(O$_{1-x}$F$_x$)FeAs superconductors even though they have a rather
high superconducting temperature at 26 K, 38 K, and 48 K for R=La,
Ce and Pr, respectively (Fig. 2). This may be related to either
strong disorder, or unconventional superconducting pairing in these
samples. The second is the ubiquitous existence of a clear Fermi
cutoff that shows little leading-edge shift in the superconducting
state (Fig. 2). Since the Fe-based compounds have multiple hole-like
Fermi surface sheets around $\Gamma$ point and electron-like sheets
around M($\pi$, $\pi$) point(inset of Fig. 2)\cite{DongSDW}, they
may have different superconducting order parameters among different
Fermi surface sheets. If the clear Fermi cut-off is not due to
non-superconducting metallic impurities in the samples, it indicates
that there are ungapped Fermi surface sheet(s) in the Fe-based
superconductors. As the laser photoemission (hv=6.994 eV) can only
cover the Fermi surface sheets around $\Gamma$
point\cite{WorkFNote}, but not those around the M($\pi$,$\pi$) point
(inset of Fig. 2), these laser data may further suggest that there
are ungapped Fermi surface sheet(s) near the $\Gamma$ point.

To identify superconducting gaps on all the Fermi surface sheets,
one has to look for signatures in the momentum-integrated
photoemission spectra at higher binding energies away from the Fermi
cutoff. It is natural to see whether the 13$\sim$16 meV feature may
represent a superconducting gap structure (Fig. 2). However, the
observation of the same feature in undoped non-superconducting
LaOFeAs and SmOFeAs\cite{HYLiu} has unambiguously ruled out this
possibility. No other clear features between the Fermi level and
13$\sim$16 meV can be identified from the laser data in
LaO$_{0.92}$F$_{0.08}$FeAs and CeO$_{0.88}$F$_{0.12}$FeAs but there
seems to be a feature near 9 meV in PrO$_{0.89}$F$_{0.11}$FeAs which
needs to be further checked on its reproducibility with higher
precision and data quality(Fig. 2).  In order to be able to probe
the superconducting gap on all the Fermi surface sheets including
those near the M($\pi$,$\pi$) point (inset of Fig. 2), we also took
high resolution data on the PrO$_{0.89}$F$_{0.11}$FeAs
superconductor at different temperatures using the Helium lamp (Fig.
1c). Like the laser data, there is no superconducting coherence
peak(s) developed and there is a clear Fermi cut-off with nearly
zero leading-edge shift below the superconducting temperature. For
the same reason as in the laser case, one has to look for the
signature of the superconducting gaps at higher binding energy but
no signatures seem to be clearly identifiable in the data except for
the 13$\sim$16 meV feature (inset of Fig. 1). We note that the
difficulty to clearly identify superconducting gaps on all the Fermi
surface sheets does not mean there are no superconducting gaps
opening in these Fe-based superconductors. It indicates further
measurements with better resolution and higher data statistics are
needed to search for the subtle features in the photoemission
spectra.

Fig. 3 shows detailed temperature dependence of the photoemission
spectra for the undoped LaOFeAs sample measured using both laser and
helium lamp.  In both cases, the temperature-induced change is
mainly confined near the Fermi level [-70meV,70meV] range; the high
binding energy spectra at different temperatures can be normalized
to overlap with each other.  Note that the spectra at different
temperatures do not cross the same point at the Fermi level, a
behavior that is different from a normal metal like gold where all
spectra cross at the same energy E$_F$. To remove the effect of
thermal broadening effect, the spectra are divided by Fermi-Dirac
distribution function at the respective temperature and shown in the
insets of Fig. 3. There is a clear suppression of the spectral
weight near the Fermi level with decreasing temperature in the laser
data (upright inset of Fig. 3a), a behavior that starts at
temperatures even as high as 150 K. The helium lamp data
(bottom-left inset of Fig. 3a) are consistent with the laser data
although the magnitude of near-E$_F$ spectral weight suppression
appears to be weaker. This spectral weight loss is similar to the
normal state spectral weight depletion near the antinodal region in
the underdoped cuprate superconductors which is related to the
opening of a pseudogap\cite{PseudogapReview}.

To further examine the possible opening of the pseudogap in Fe-based
compounds, we follow the procedure that is commonly used in
high-T$_c$ cuprate superconductors\cite{NormanSymmetrize} by
symmetrizing the original data in Fig. 3a with respect to the Fermi
level (Fig. 3b). This is another way to remove the Fermi-Dirac
distribution function and it provides a visualized way to look for a
gap. Again one sees clearly the depletion of spectral weight near
the Fermi level with decreasing temperature (Fig. 3b).  To highlight
the effect caused by temperature, we further divide the symmetrized
spectra with the one at 200 K (Fig. 3c).  The suppression of the
spectral weight near the Fermi level becomes clearer and one can now
identify an energy scale at which the spectral weight starts to
lose. It is in the 25$\sim$40 meV energy range for different
temperatures (Fig. 3c).

The temperature dependence of photoemission spectra for the
superconducting samples (Fig. 4) appears to be surprisingly similar
among different materials and to that in the undoped LaOFeAs sample
(Fig. 3). Here again one sees suppression of spectral weight with
decreasing temperature as in the inset of Fig. 4a, symmetrized data
in Fig. 4b1, c1 and d1 and normalized data in Fig. 4b2, c2 and d2
for LaO$_{0.92}$F$_{0.08}$FeAs, CeO$_{0.88}$F$_{0.12}$FeAs and
PrO$_{0.89}$F$_{0.11}$FeAs samples, respectively. We note that our
results on the LaO$_{0.92}$F$_{0.08}$FeAs sample are rather
different from that measured by Ishida et al.\cite{Shin}.
Particularly, we do not observe any signature of 0.1 eV pseudogap as
claimed by them\cite{Shin}. This difference may be caused by sample
surface cleanness because we found that the 0.2 eV peak feature in
our sample (Fig. 1) is much sharper than that in theirs\cite{Shin}.
For these superconducting samples with T$_c$=26$\sim$48 K, one may
wonder whether the near-E$_F$ spectral weight suppression in the
superconducting state, like the 13 K  data in Fig. 4, compared with
a normal state data at 50 K, can be taken as a signature of
superconducting gap opening, as did by Sato et al.\cite{Takahashi}.
Noting that the same behavior also occurs in the non-superconducting
LaOFeAs sample (Fig. 3), we believe this is not a reliable way in
judging on a superconducting gap.

In summary, from our extensive high resolution photoemission
measurements on the R(O$_{1-x}$F$_x$)FeAs (R=La,Ce and Pr)
compounds, together with previous measurements on
Sm(O$_{1-x}$F$_x$)FeAs\cite{HYLiu}, we have identified several
universal features in the electronic structure of the Fe-based
compounds. These universal features can provide important
information about band structure, superconducting gap and pseudogap
in these Fe-based compounds. The parent LaOFeAs compound shows a
metallic nature that is distinct from the parent compound of cuprate
superconductors that is a Mott insulator. The origin of the
13$\sim$16 meV feature and whether it can be due to electron
coupling with some collective excitations need to be further
studied. The zero leading-edge shift in the spectra in the
superconducting state suggests that the Fermi surface sheet(s)
around the $\Gamma$ point may not be gapped. The spectral weight
suppression near E$_F$ with decreasing temperature points to
possible existence of a pseudogap in the Fe-based compounds. Its
origin as to whether it can be caused by local SDW fluctuation or
strong electron-boson coupling needs further experimental and
theoretical studies.

This work is supported by the NSFC, the MOST of China (973 project
No: 2006CB601002, 2006CB921302), and CAS (Projects ITSNEM).

$^{*}$Corresponding author: XJZhou@aphy.iphy.ac.cn



\begin{thebibliography}{99}
\bibitem{VBNote}Because of the relatively large spot size of the helium lamp light,
the high binding energy spectra may contain weak signals from
graphite surrounding the sample, but the low binding energy part
within 1 eV is not affected.
\bibitem{NotePosition}The uncertainty of the position determination is
$\pm$3 meV.
\bibitem{WorkFNote}The work function is assumed to be 4.3 eV.
\end{thebibliography}
\end{document}